\documentclass{appolb}
\usepackage{epsfig}
\usepackage{graphicx}

\newcommand{\rts}{\mbox{$\sqrt{\mathrm{s}_{_{\mathrm{NN}}}}$}}

\newcommand{\DDbar}{${\rm D}\overline{\rm D}$}

\begin{document}
\title{Bulk Properties of QCD-matter at Highest Collider Energies
}
\author{K. Schweda
\address{Physikalisches Institut, Universit\"at Heidelberg, Philosophenweg 12, \\ D-69120 Heidelberg, Germany}
}
\maketitle
\begin{abstract}
The Large Hadron Collider at CERN will provide Pb-Pb collisions at energies up to \rts = 5.5~TeV. We speculate on global observables, i.e. the charged particle density at mid-rapidity, chemical freeze-out conditions and collective parameters for transverse radial an elliptic flow. Finally, we present an idea how to address the important issue of thermalization by measuring the correlated production of heavy-quark hadrons.    
\end{abstract}
  
\section{Introduction}
Lattice QCD calculations, at vanishing or finite net-baryon density,
predict a cross-over transition from the deconfined thermalized partonic
matter -- the quark-gluon plasma -- to hadronic matter at a critical
temperature $T_{\rm c} \approx 150$--180~MeV~\cite{karsch}.
The Large Hadron Collider at CERN will provide nuclear collisions at unprecedented high energies. 
The huge increase in collision energy should allow for the unambiguous identification and characterization of the quark-gluon plasma.      
In these proceedings we present a few topics, among a variety of interesting probes available at LHC.

\section{Charged-particle multiplicity}

\begin{figure}[t]
\begin{minipage}{0.49\textwidth}
\begin{center}
\vspace{-2.8cm}
\includegraphics[width=0.8\textwidth]{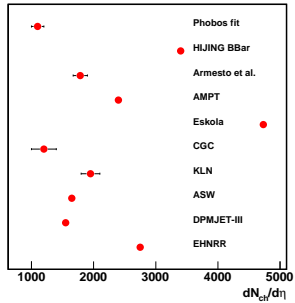}
\end{center}
\end{minipage}
\begin{minipage}{0.49\textwidth}
\begin{center}
\caption{(Color online) Predictions of mid-rapidity charged-particle densities from empirical extrapolations and various model calculations. These numbers have been taken from~\cite{hi_CERN07}.} 
\label{fig1}
\end{center}
\end{minipage}
\vspace{-0.4cm}
\end{figure}
%
%
One of the first observables measured at LHC will be the mid-rapidity density of charged particles. Recent predictions~\cite{hi_CERN07} from empirical extrapolations and model calculations are summarized in Fig.~\ref{fig1}. Most numbers fall between one thousand to three thousand charged particles per unit rapidity. 
The charged-particle density can be used to estimate the early-stage energy density of the system, assuming the system is thermalized. Even moderate numbers on mid-rapidity densities lead to an energy density and corresponding temperature several times above the critical values resulting in a long life time of the quark-gluon plasma.    
\section{Chemical Equilibrium}
A crucial question to be addressed in high-energy nuclear collisions is the degree of thermalization. 
Hadron yields, or their ratios respectively, are well described within the Statistical Model over a large range of energies from AGS, SPS and RHIC assuming a specific equilibrium temperature and baryon chemical potential~\cite{anton_2006}. The chemical potential decreases smoothly with increasing center-of-mass collision energy, i.e. new baryons and anti-baryons are created with increasing ease. At LHC energies, the baryon chemical potential might be as low as $\mu_{\rm B}$ = 1 MeV. As a consequence, the vast majority of baryons and anti-baryons is generated in pair-creation-like processes. Baryon-number transport from large beam-rapidity to mid-rapidity contributes little to the production of baryons at mid-rapidity. Hence, anti-baryon to baryon ratios approach unity. 
To quantitatively address baryon-number transport at LHC energies, a precision in the proton and anti-proton yield of better than 1\% is essential. This imposes a formidable challenge at LHC, given the considerably large experimental background of low-energy protons at a collider.      

The chemical freeze-out temperature increases strongly at lower collision energies and then plateaus rather abruptly near \rts = 10 GeV at values around $T_{\rm ch}$ = 160 MeV. This value coincides with early predictions  for the critical temperature from lattice gauge QCD calculations as well as with the limiting temperature at which hadrons can live, as derived by Hagedorn~\cite{hagedorn_65}. These findings support the idea that indeed the boundary between hadrons as the relevant degree of freedom on the one side and quarks and gluons on the other side is reached at a critical collision energy. Additional energy above this critical energy goes into further heating which in turn leads to prolonged cooling and subsequent freeze-out into hadrons at the boundary.            

\section{Partonic Collectivity}
\begin{figure}[htb]
\begin{minipage}{0.49\textwidth}
\begin{center}
\vspace{-3.6cm}
\includegraphics[width=1.0\textwidth]{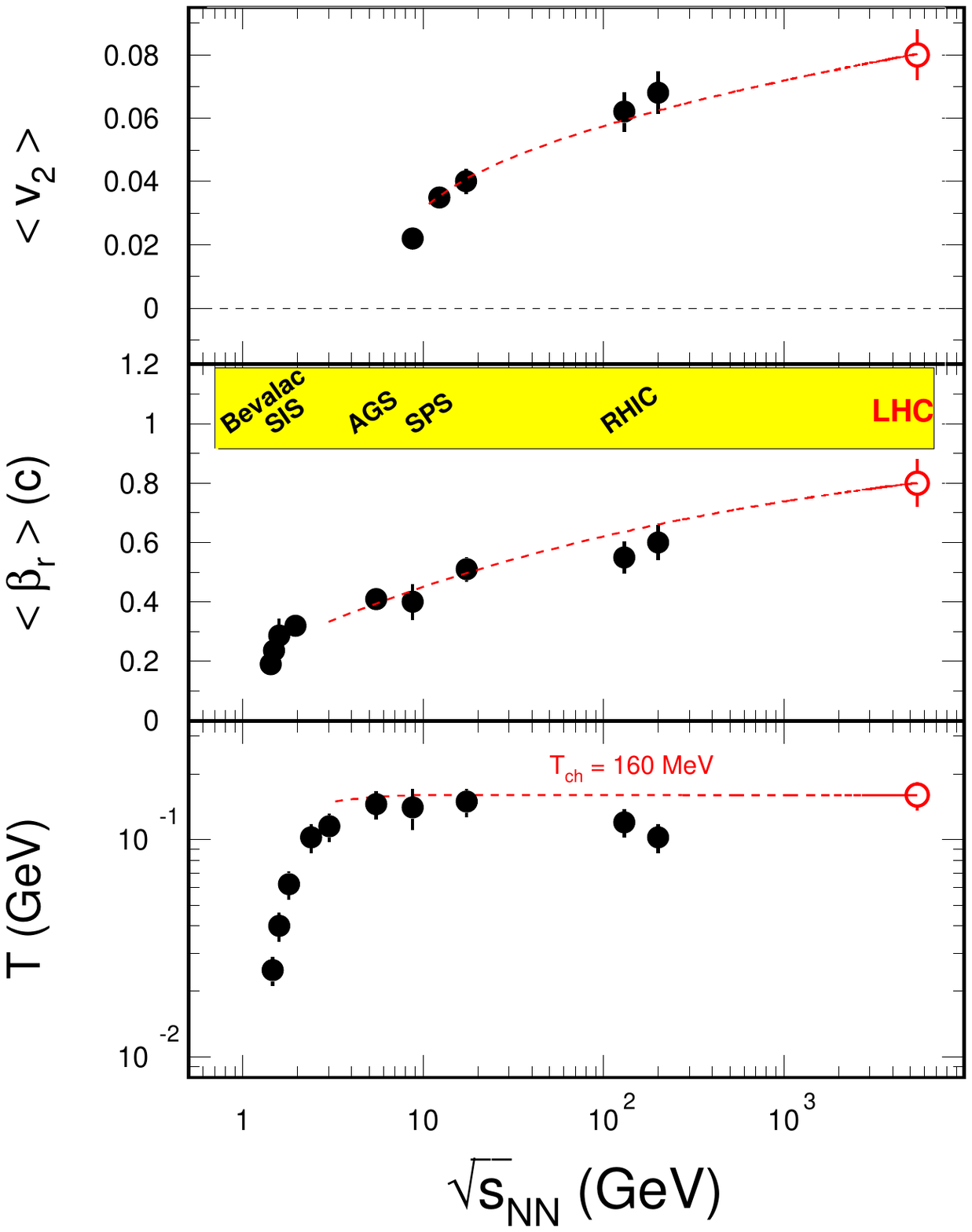}
\end{center}
\end{minipage}
\begin{minipage}{0.49\textwidth}
\begin{center}
\caption{(Color online) Experimental data~(closed circles) on integrated elliptic flow parameters $v_2$ for minimum bias collisions~(top), integrated collective flow velocity for central collisions~(middle), and corresponding kinetic freeze-out temperature parameter $T_{\rm fo}$ as a function of center-of-mass collision energy. The points at LHC energies~(filled circles) are our predictions from simple extrapolation.} 
\label{fig3}
\end{center}
\end{minipage}
\vspace{-0.4cm}
\end{figure}
If local thermal equilibrium is reached in the system, the thermodynamic pressure and global pressure gradients will drive the system to develop hydrodynamic-like collective flow.
A characteristic feature of hydrodynamic flow is the mass ordering in the anisotropy parameter at low momentum, with larger values of the anisotropy parameter for the light pions to lower values until the heavy $\Omega$-baryon as observed at RHIC energies~\cite{STAR_v2_2005}. 

The evolution of collectivity parameters of pions, kaons and protons as a function of collision energy is summarized in Fig.~\ref{fig3}. At lower collision energies, the integrated $v_2$ and mean transverse flow velocity $\beta$ rise rapidly. At higher energies, they still increase monotonically but less rapidly.
A similar trend is observed for the kinetic freeze-out temperature parameter $T_{\rm fo}$, it seems even to saturate at intermediate collision energies. These trends suggest a softening of the equation of state at intermediate collisions energies. The most violent collisions occur at RHIC energies with values of $v_2$ = 7\% and $\beta$ = 60\%  (in units of the speed of light).
At LHC energies, due to the larger initial temperatures and energy densities, even larger values on $v_2$ and $\beta$ are expected. 
To estimate expectations on $v_2$ and $\beta$ for LHC energies, we simply use a fit to the experimental data by a square root function and a $(1-\exp)$ function, respectively. Extrapolation to LHC energies leads to $v_2$ = 8\% and $\beta$ = 80\%.  Interestingly, these values are identical to those predicted by hydrodynamical calculations~\cite{hi_CERN07}.
As argued earlier, $T=$160~MeV is the limiting temperature for hadrons. The large collective flow will lead to fast dilution of the system immediately after hadronization, practically leading to an instantaneous kinetic freeze-out. Hence, we expect the kinetic freeze-out temperature parameter $T_{\rm fo}$ coinciding with the limiting temperature of 160~MeV. 
However, spectra from high-energy $p+p$ collisions are also best described by a similar temperature parameter. This leads to two extreme dynamical scenarios at LHC with identical kinetic freeze-out temperature parameters. On the one hand formation of a long-lived quark-gluon plasma with a collective flow velocity of 80\%. On the other hand a simple superposition of $p+p$ collisions with an effective flow velocity $\beta$ = 30\%, similar to those observed in $p+p$ collisions at RHIC.     
\begin{figure}[tb]
\begin{minipage}{0.49\textwidth}
\begin{center}
\includegraphics[width=1.0\textwidth]{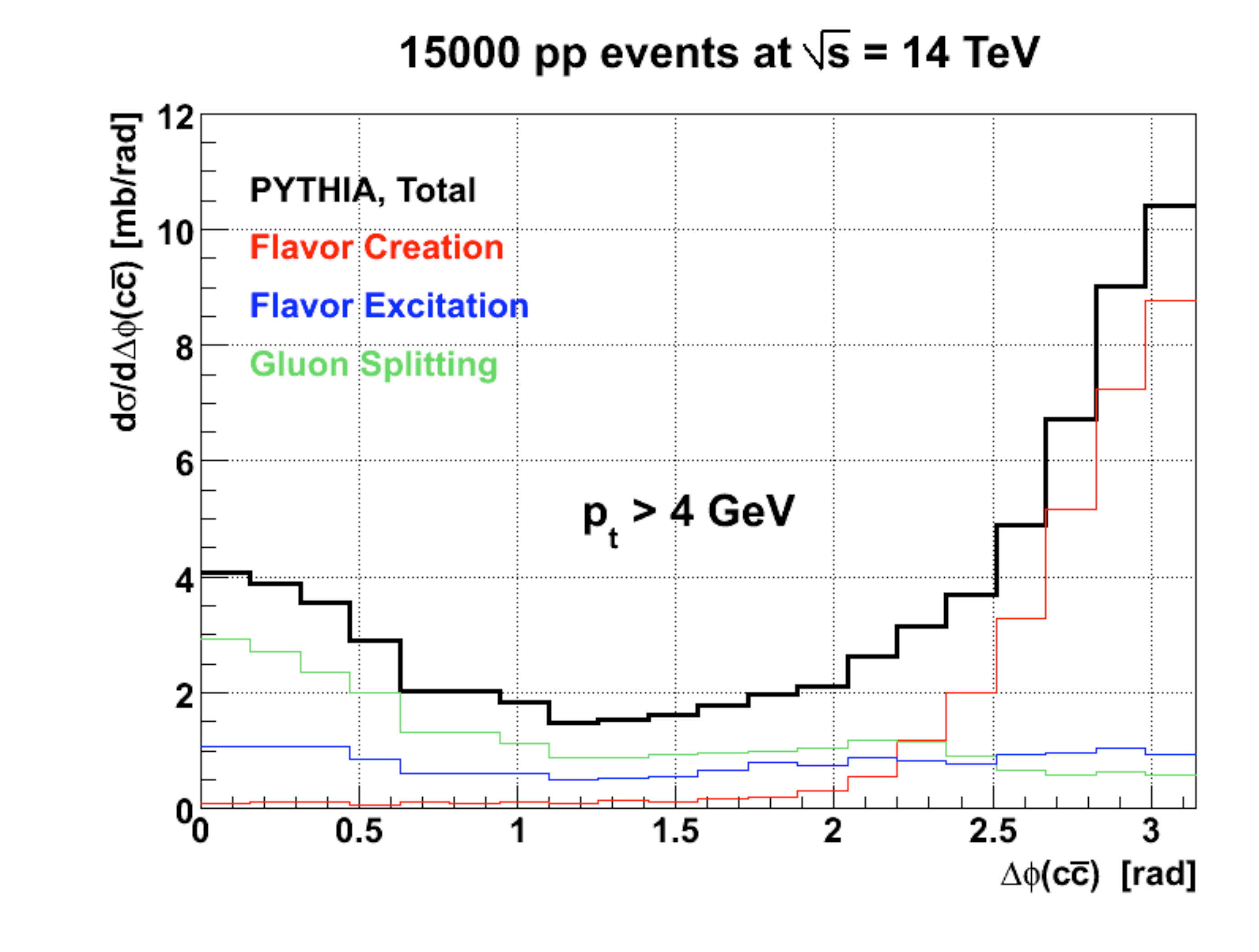}
\end{center}
\end{minipage}
\begin{minipage}{0.49\textwidth}
\begin{center}
\caption{(Color online) Relative azimuthal correlations of \DDbar\ pairs from $p+p$ collisions at \rts = 14~TeV as calculated with PYTHIA. The thick line is the sum of contributions from gluons splitting, flavor excitation, and pair creation.} 
\label{fig4}
\end{center}
\end{minipage}
\vspace{-0.4cm}
\end{figure}

Recent predictions on anisotropy parameters at LHC energies~\cite{hi_CERN07} have stimulated some excitement. When compared to RHIC energies at the same momentum, values for pions increase while they decrease for the heavier protons. However, this trend is expected from hydrodynamic behaviour. At LHC energies, the higher initial temperature results in a larger collective flow velocity, i.e. protons are pushed to larger momentum. In contrast, due to their small mass, pions are less affected by flow. Their momentum distribution is dominated by the freeze-out temperature. Hence, the overall increase in transverse radial and elliptic flow moves anisotropy parameters for pions up, while the values for protons move up {\it and} to larger momentum. The strong linear increase of $v_2$ at low momentum then leads to an apparent decrease for protons when compared to RHIC energies at the same momentum.               

Heavy-quark hadrons have been suggested as a probe of thermalization at the partonic stage~\cite{schweda_2006}.  The production of a heavy-quark is accompanied by its anti-quark and thus correlated. This initial correlation survives the hadronization process and hence is observable in the azimuthal correlation of e.g. \DDbar\ hadrons, as shown in Fig.~\ref{fig4}. At LHC energies, three mechanisms contribute significantly to the production of charm quarks, (a) gluon splitting resulting in a forward correlation at small relative azimuthal angles $\Delta \phi$, (b) flavor excitation exhibiting rather flat correlations and (c) pair creation leading to a backward correlation around $\Delta \phi = \pi$. Strong collective flow in $Pb-Pb$ collisions will generate additional forward correlations. Due to their large mass, heavy-quark hadrons are less subject to hadronic interactions. Thus modifications of these initial correlations can be dominantly attributed to interactions at the partonic stage. A weakening or even complete absence of backward correlations in \DDbar\ correlations thus indicates frequent interactions at the partonic stage. Hence, at least light quarks (u,d,s) are likely to be in thermal equilibrium. 

If partonic interactions occur frequently enough, heavy-quarks reach kinetic equilibrium. This lead to the idea of statistical hadronization of charm quarks~\cite{pbm}. Since charm quarks are abundantly produced at LHC energies, they might recombine leading to an increase in $J/\psi$ production by a factor two compared to $p+p$ collisions~\cite{pbm_2007}. Furthermore, the production of rarely produced hadrons with multiple heavy-quarks, e.g. $\Xi_{\rm cc}$,  $\Omega_{\rm cc}$, and mixed heavy-quarks, e.g $B_{\rm c}$ might be enhanced by an order of magnitude~\cite{beccatini_2005}. 


\section{Conclusions}
The LHC startup is scheduled for early summer 2008 providing $p+p$ collisions at top energies. Roughly a year later, $Pb-Pb$ collisions will become available at unprecedented high energies, a factor 30 higher than presently available. The quark-gluon plasma expected to be created in these collisions will provide initial energy densities and temperatures well above the critical conditions. The mid-rapidity charged particle density is likely to be between one and three thousand. Stable hadrons will be exclusively produced at the phase boundary, at $T_{ch}$ = 160~MeV with a chemical potential of $\mu_B$ less than 5~MeV. We expect the development of strong collective flow at the partonic stage with parameters for integrated $v_2$ = 8\% in minimum bias collisions and average flow velocity $\beta$ = 80\% in central collisions with a kinetic freeze-out temperature of $T_{\rm fo}$ = 160~MeV, coinciding with the limiting temperature for hadrons. This together with the large abundance of heavy-quarks makes LHC a unique laboratory to study QCD matter at extreme densities and temperatures.

\end{document}